\documentstyle[12pt,epsfig,a4]{article} 
\newcommand{\vs}{\vspace{-0.25cm}}
\begin{document}
\begin{center}
{\Large{\bf Resummation of in-medium ladder diagrams:\\ s-wave effective 
range and p-wave interaction}\footnote{Work supported in part by DFG and 
NSFC (CRC 110).}}
\bigskip

N. Kaiser\\
\medskip
{\small Physik-Department T39, Technische Universit\"{a}t M\"{u}nchen,
    D-85747 Garching, Germany

\smallskip
{\it email: nkaiser@ph.tum.de}}
\end{center}
\medskip
\begin{abstract}
A recent work on the resummation of fermionic in-medium ladder diagrams to 
all orders is extended by considering the effective range correction in the
s-wave interaction and a (spin-independent) p-wave contact-interaction. A 
two-component recursion generates the in-medium T-matrix at any order 
when off-shell terms spoil the factorization of multi-loop diagrams. 
The resummation to all orders is achieved in the form of a geometrical series for 
the particle-particle ladders, and through an arctangent-function for the 
combined particle-particle and hole-hole ladders. One finds that the  
effective range correction changes the results in the limit of large 
scattering length considerably, with the effect that the Bertsch parameter $\xi_n$
nearly doubles. Applications to the equation of state of neutron matter at low 
density are also discussed. For the p-wave contact-interaction the resummation 
to all orders is facilitated by decomposing tensorial loop-integrals with a 
transversal and a longitudinal projector. The enhanced attraction provided by  
the p-wave ladder series has its origin mainly in the coherent sum of Hartree and 
Fock contributions. 
\end{abstract}

\bigskip

\vspace{-0.3cm}
\section{Introduction and summary}
Dilute degenerate many-fermion systems with large scattering lengths are of 
interest, e.g. for modeling the low-density behavior of nuclear and neutron 
star matter. Because of the possibility to tune magnetically atomic interactions, 
ultracold fermionic gases provide an exceptionally valuable tool to explore the 
non-perturbative many-body dynamics involved in the crossover from the 
superconducting to the Bose-Einstein condensed state (for a recent comprehensive 
review of this fascinating field, see ref.\cite{zwerger}). Of particular interest 
in this context is the so-called unitary limit in which the two-body interaction 
has the critical strength to support a bound-state at zero energy. As a consequence 
of the diverging scattering length, $a\to \infty$, the strongly interacting 
many-fermion system becomes scale-invariant. Its ground-state energy is then determined 
by a single universal number, the so-called Bertsch parameter $\xi$, which 
measures the ratio of the energy per particle $\bar E(k_f)^{(\infty)}$ to that of 
a free Fermi gas $\bar E(k_f)^{(0)}=3k_f^2/10M$. Here, $k_f$ denotes the Fermi 
momentum and $M$ the fermion mass.

The calculation of $\xi$ is an intrinsically non-perturbative problem which has
been approached by numerical quantum Monte Carlo simulations. As the state of 
the art, a value of $\xi\simeq 0.38$ emerges at present \cite{montecarlo,effrange}.
It is consistent with a recent precision measurement of the equation of state by 
the MIT group, which gives $\xi=0.37\pm 0.01$ \cite{zwierlein} (extrapolating to
zero temperature). Among the analytical approaches let us mention the 
$\epsilon$-expansion by Nishida and Son \cite{epsilon}, where $\epsilon=4-d$ with 
$d$ the number of space dimensions. A naive extrapolation to $d=3$ gives 
$\xi=0.475$, but with inclusion of higher order corrections and Pade 
interpolations between the (otherwise diverging) expansions around $d=2$ and 
$d=4$, a good value of $\xi = 0.38\pm0.01$ could be obtained (see chapter 7 in 
ref.\cite{zwerger}). Because of the very large neutron-neutron scattering length 
$a_{nn}\simeq 19\,$fm, neutron matter at low densities is supposed to be a 
Fermi gas close to the unitary limit. Quantum Monte Carlo simulation 
\cite{gandolfi,neutronqm} and other sophisticated many-body calculations  
of neutron matter give indications for a value of $\xi_n \simeq 0.5$.    

In a recent work \cite{resum} the complete resummation of the (combined) 
particle-particle and hole-hole ladder diagrams generated by a 
contact-interaction proportional to the scattering length $a$ has been achieved.  
A key to the solution of this (restricted) problem has been a different 
organization of the many-body calculation from the start. Instead of treating 
(propagating) particles and holes separately, these are kept together and the 
difference to the propagation in vacuum is measured by a ``medium-insertion''.  
The following identical rewriting of the (non-relativistic) particle-hole 
propagator:
\begin{eqnarray} 
G(p_0,\vec p\,) &=&  {i\, \theta(|\vec p\,|-k_f)\over p_0-
\vec p^{\,2}/2M+i \epsilon }+{i\, \theta(k_f-|\vec p\,|)\over p_0-\vec p^{\,2}/2M
-i \epsilon} \nonumber \\ &=& {i \over p_0-\vec p^{\,2}/2M+i \epsilon } 
-2\pi\, \delta(p_0-\vec p^{\,2}/2M)\,\theta(k_f-|\vec p\,|)\,, \end{eqnarray}
explains the basic principle of the approach. In that organizational scheme the 
pertinent in-medium loop  is complex-valued, and therefore the contribution to 
the energy per particle $\bar E(k_f)$ at order $a^n$ is not directly obtained 
from the $(n-1)$-th power of the in-medium loop (see Fig.\,1). However, after 
reinstalling the symmetry factors $1/(j+1)$  which belong to diagrams with $j+1$ 
double medium-insertions, a real-valued expression is obtained for all orders 
$a^n$. Known results about the low-density expansion \cite{hammer,steele} up to and 
including order $a^4$ could be reproduced with improved numerical accuracy. The 
emerging series in $a k_f$ could even be summed to all orders in the form of a 
double-integral over an arctangent-function. In that explicit representation the 
unitary limit, $a\to \infty$, could be taken straightforwardly and the value 
$\xi=0.507$ has been found for the Bertsch parameter \cite{resum}. This number is 
to be compared with the value $\xi^{(pp)}\simeq 0.237$ resulting from a resummation 
of particle-particle ladder diagrams only \cite{schaefer}. Clearly, since only the 
(infinite) subset of particle-hole ladder diagrams \cite{resum} is computed in the 
normal phase one cannot expect accurate results for the unitary Fermi gas, whose 
true ground state is a superfluid. But interestingly, extrapolations of the 
equation of state of the unitary Fermi gas from finite to zero temperature, which 
smoothly pass over the pairing transition at $T_c$ give indications for a Bertsch 
parameter $\xi_n \simeq 0.5$ in the normal phase \cite{zwerger,zwierlein,
discusszwerger}. In comparison to this synthetic result, the value $\xi_n=0.507$ of 
ref.\cite{resum} appears to be a rather good finding.    
 
The purpose of the present paper is to extend the resummation method of 
ref.\cite{resum} by including the effective range correction in the s-wave 
interaction and by considering also  a spin-independent p-wave contact 
interaction to all orders. The effective range correction to a contact-interaction
(proportional to the scattering length $a$) has been studied first by Sch\"afer et al. 
in ref.\cite{schaefer} for the resummed particle-particle ladder series. The 
implications of this correction on the equation of state have, however, not been 
explored in much detail and only rough perturbative estimates have been given. In a 
related work by Schwenk and Pethick \cite{schwenk} it has been argued that the 
inclusion of the effective range $r_s=2.75\,$fm leads to an improved description 
of neutron matter at low densities. Their implementation of the effective range $r_s$ 
into the particle-particle ladder series assumes a form similar to the effective 
range expansion for neutron-neutron scattering in free space. However, a proper 
diagrammatic calculation \cite{schaefer} shows that in the medium the combination 
of scattering length $a$ and effective range $r_s$  produces additional 
(density-dependent) off-shell terms which spoil the factorization of multi-loop 
diagrams. A resummation of the particle-particle ladder diagrams to all orders  
in the form of a geometrical series is still possible, but it differs from 
the expression suggested by the effective range expansion in vacuum.

The present paper is organized as follows: In section 2 we perform the resummation 
of fermionic in-medium ladder diagrams for a contact-interaction that includes the 
effective range $r_s$. It is demonstrated that a two-component recursion generates 
the in-medium T-matrix at any order when off-shell terms spoils the direct 
factorization of multi-loop diagrams. The resummation to all orders is still achieved  
in the form of a geometrical series for the particle-particle ladders, and 
through an arctangent-function for the combined particle-particle and hole-hole 
ladders. The second more complicated case involves yet a (combinatorial) 
conjecture because the arctangent-series could be verified so far with diagrammatic 
methods only to some finite order. One finds that the effective range correction 
changes the results in the limit of large scattering length $a\to \infty$ 
considerably, with the effect that the Bertsch parameter $\xi_n$ (for the normal 
phase) increases. For the combined particle-hole ladder series one gets $\xi_n
(r_s\ne 0) = 0.876$, and for the resummed particle-particle ladder series alone 
the value is $\xi_n^{(pp)}(r_s\ne 0)\simeq 0.43$. Applications of these
resummation results to the equation of state of neutron matter 
at low density are presented and discussed in comparison. Section 3 deals with the 
analogous resummation to all orders for a (spin-independent) p-wave 
contact-interaction. In this case the factorization of multi-loop diagrams is 
achieved more directly by decomposing tensorial loop-integrals into a transversal 
and a longitudinal part. These two parts decouple and can therefore be resummed 
separately through an arctangent-function or a geometrical series. The enhanced 
attraction provided by the p-wave ladder series (in the limit of large scattering 
volume $a_1^3\to \infty$) has its origin mainly in the coherent sum of Hartree 
and Fock contributions. In the appendix analytical results for the resummation of 
particle-hole diagrams to all orders are presented for a two-component fermionic 
system  (such as nuclear matter) with two different scattering lengths, $a_s$ 
and $a_t$, in the presence of an isospin-asymmetry, $k_n>k_p$. A special case of 
such an asymmetric system is the (unitary) Fermi gas in the spin-unbalanced 
configuration.    

Presumably the results of the present paper are contained 
(to a certain extent) as special cases in the work of Lacour et al. \cite{lacour}. In 
that paper non-perturbative methods for chiral effective field theory of finite density 
nuclear systems have been developed. The approach is based on a partial wave 
representation of leading order nucleon-nucleon interactions which are summed to all orders 
in the medium.
A detailed comparison of the present results (obtained from simple contact-interactions) 
with those of ref.\cite{lacour} is certainly interesting, but this requires an extensive 
communication between both groups in order to translate the different formalisms 
correctly into each other and to establish the common range of applicability.  
 
The study of strongly-coupled fermionic many-body systems (such as neutron 
matter) at low density with quantum Monte Carlo simulations \cite{spinpolarn,
gezsharma,superfluid} is presently an active research area. The analytical 
results of the present work, which of course cover only a restricted part of 
the many-body dynamics, could be potentially useful for interpreting these 
numerical simulations.     

\section{Resummation with inclusion of the effective range}
\begin{figure}
\begin{center}
\includegraphics[scale=0.6]{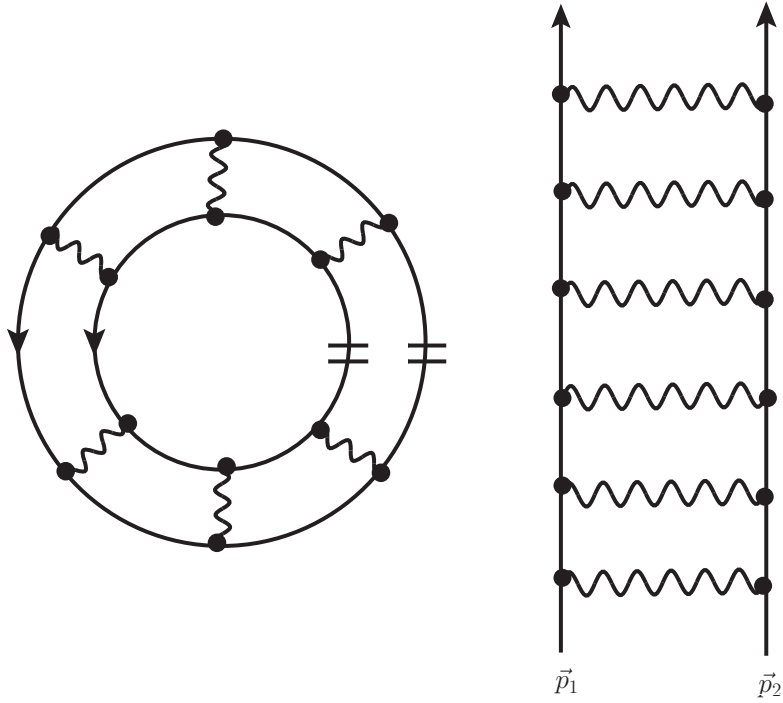}
\end{center}
\vspace{-.6cm}
\caption{Left: Closed multi-loop diagram representing a contribution to the 
energy density. Right: Planar ladder diagram obtained by opening at a minimal 
pair of adjacent medium-insertions. Wiggly lines symbolize contact-interactions.}
\end{figure} 

\begin{figure}
\begin{center}
\includegraphics[scale=0.6]{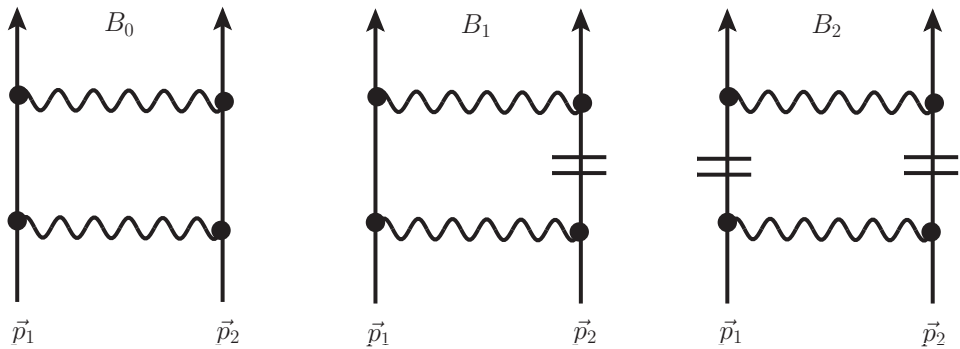}
\end{center}
\vspace{-.6cm}
\caption{In-medium loop organized in the number of medium-insertions ({\bf =}). The 
central diagram has a reflected partner. The external momenta $|\vec p_{1,2}|<k_f$ are
inside the Fermi sphere.}
\end{figure}

In this section we develop methods which allow for a complete resummation of 
fermionic in-medium ladder diagrams for an s-wave contact-interaction that 
includes the effective range correction. We recall from ref.\cite{resum} that 
after opening a closed multi-loop diagram (representing the energy density) at 
adjacent medium-insertions a planar ladder diagrams results (see Fig.\,1). If the 
contact-interaction is momentum-independent all these (open) multi-loop diagrams 
factorize into powers of the in-medium loop. In the ordering scheme implied by 
eq.(1) the in-medium loop is composed of contributions with zero, one, and two 
medium-insertions (see Fig.\,2). The first part represents the rescattering in the 
vacuum and the other two parts incorporate systematically the (Pauli-blocking) 
effects due to the fermionic medium. The resulting in-medium loop is 
complex-valued, but after a careful combinatoric analysis of the closed multi-loop 
diagrams a real-valued expression for the energy density is obtained at each 
order. In the next step the complete resummation to all orders can be performed 
through an arctangent-function that depends on both the real and imaginary part 
of the in-medium loop.  
      
The aim is to generalize this novel approach to an s-wave contact-vertex: 
\begin{equation} C_0+C_2 (\vec q_{\rm in}^{\,2}+\vec q_{\rm out}^{\,2})\,,
\end{equation} 
which includes the effective range correction. Here, $\vec q_{\rm in}$ and $\vec 
q_{\rm out}$ denote half of the momentum difference in the initial and the final 
state, respectively. Such a form of the contact-vertex is actually dictated by 
Galilei invariance. The relations of the coupling constants $C_0$ and $C_2$ to the 
scattering length $a$ and the effective range parameter $r_s$ are: 
\begin{equation} C_0 = {4\pi a\over M}\,, \qquad  C_2 = -{\pi a^2 r_s\over M}
\,.\end{equation} 
Note that we are choosing the sign-convention such that a positive scattering 
length $a>0$ corresponds to attraction. The repeated rescatterings via the 
contact-vertex eq.(2) in vacuum (i.e. the left diagram in Fig.\,2 iterated to 
all orders) leads to an s-wave phase shift $\delta_0(q)$ of the form: 
\begin{equation}\tan \delta_0(q) = a\, q-{a^2 r_s\over 2} q^3 \,,\end{equation} 
with $q$ the center-of-mass momentum. We are using dimensional regularization, 
where all (scale-free) power divergences are set to zero: $\int_0^\infty dl\,l^n 
= 0$. The real part of the in-medium loop is generated by the central diagram 
in Fig.\,2. The pertinent integration formula for the real part reads:
\begin{equation} -\!\!\!\!\!\!\int\!{d^3 l \over (2\pi)^3}\, {-M \over \vec l^{\,2}- 
\vec q^{\,2}} \Big\{\theta(k_f-|\vec P-\vec l\,|)+ \theta(k_f-|\vec P+
\vec l\,|)\Big\}\dots\,,\end{equation} 
with the factors from the lower and upper interaction-vertex to be inserted at 
the place of the dots. Here, we have introduced the half sum $\vec P = (\vec p_1
+\vec p_2)/2$ and half difference $\vec q = (\vec p_1-\vec p_2)/2$ of the external 
momenta $|\vec p_{1,2}|<k_f$ inside the Fermi sphere. All three diagrams in Fig.\,2 
contribute to  imaginary part of the in-medium loop and the pertinent 
integration formula reads: 
\begin{eqnarray} \int\!\!{d^3 l \over (2\pi)^3} \,M\pi\, \delta(\vec l^{\,2}-
\vec q^{\,2}) \,\theta(k_f-|\vec P-\vec l\,|)\, \theta(k_f-|\vec P+\vec l\,|) 
\dots \,.\end{eqnarray}
Note that we have left out a term $[1-\theta(\dots)][1-\theta(\dots)]$ which 
vanishes on-shell $(\vec l^{\,2}=\vec q^{\,2})$ due to energy conservation and 
Pauli-blocking (see eq.(8) in ref.\cite{resum}). Let us now analyze more closely 
the one-loop diagrams in Fig.\,2 with two momentum-dependent $C_{0,2}$ vertices. 
For the lower vertex the assignment $\vec q_{\rm in}=\vec q $, $ \vec q_{\rm out}
=\vec l$ with $\vec l$ the loop-momentum holds, while for the upper vertex it is 
$ \vec q_{\rm in}=\vec l$, $ \vec q_{\rm out}=\vec q$. Actually, if a further loop 
is considered the second $ \vec q_{\rm out}$ is set to the new loop momentum 
$\vec l\,'$ and therefore this second $\vec q_{\rm out}$ should not be fixed 
immediately. One recognizes that the product of interaction vertices contains 
terms proportional to $\vec l^{\,2}$ and $\vec l^{\,4}$. With inclusion of the 
energy denominator in eq.(5) these can be decomposed into on-shell vertices 
($\vec l^{\,2} \to \vec q^{\,2}$) and polynomial off-shell pieces:  
\begin{equation}  {\vec l^{\,2} \over \vec l^{\,2}- \vec q^{\,2}} = {\vec q^{\,2}
\over \vec l^{\,2}- \vec q^{\,2}}+1\,,\qquad  {\vec l^{\,4} \over \vec l^{\,2}- 
\vec q^{\,2}} = {\vec q^{\,4} \over \vec l^{\,2}- \vec q^{\,2}}+\vec l^{\,2}+ 
\vec q^{\,2}\,. \end{equation}
These off-shell terms produce additional real-valued contributions which enter 
into the next loop, where by the same mechanism new off-shell terms are 
generated, and so on. The factorization of multi-loop diagrams into powers of 
the in-medium loop gets thus spoiled by the presence of the $C_2$ coupling 
proportional to the effective range $r_s$. Although it looks complicated at first 
sight, one can recursively follow the continuous proliferation of off-shell terms 
by simultaneously keeping track of two quantities: the (on-shell) in-medium T-matrix 
$T_n$ and the coefficient $\beta_n$ of $ \vec q_{\rm out}^{\,2}$. Injecting the 
expression $T_n+\beta_n(\vec q_{\rm out}^{\,2}-\vec q^{\,2})$  at order $n$ with 
$\vec q_{\rm out}\to \vec l$ and concatenating it through the in-medium 
loop with one additional $C_{0,2}$ vertex produces as an output the same expression at 
order $n+1$. As a transparent result one deduces a linear two-component recursion of 
the form: 
\begin{equation} \left(\!\! \begin{array} {c} T_{n+1}\\ \beta_{n+1}\\ \end{array}
\!\!\right) = {\cal M} \left(\!\! \begin{array} {c} T_n \\ \beta_n  \\ 
\end{array}\!\!\right)\,, \end{equation} 
with the starting values $T_1 =C_0+2C_2 \vec q^{\,2}$ and $\beta_1 = C_2$. The 
pertinent $2\times 2$ matrix ${\cal M}$ reads:
\begin{equation} {\cal M} = \left(\!\! \begin{array} {cc} (C_0+2C_2 \vec q^{\,2})
{\cal L}+C_2 X & (C_0+C_2 \vec q^{\,2})X+C_2 Y \\ C_2 {\cal L} & C_2 X \\ 
\end{array}\!\!\right) \,,\end{equation} 
with $X,\, Y$ the off-shell terms originating from Fermi sphere integrals over the 
terms $1,\,\vec l^{\,2}$ in eq.(7) and ${\cal L}$ the complex-valued in-medium 
loop (see eqs.(14-17)). Although the polynomial expressions for the in-medium 
T-matrices $T_n$ get increasingly more complicated with increasing $n$, their 
total sum can be easily calculated with the help of matrix-inversion:  
\begin{equation} \sum\limits_{n=1}^\infty T_n = (1,0) \,\Big[{\bf 1}-{\cal M}
\Big]^{-1}\!\left(\!\! \begin{array} {c} C_0+2C_2 \vec q^{\,2} \\ C_2  \\ 
\end{array}\!\!\right)=\Big[V_{\rm eff}^{-1}- {\cal L}\Big]^{-1} =
\sum\limits_{n=1}^\infty V_{\rm eff}^{n}\, {\cal L}^{n-1} \,.\end{equation} 
The second equality reveals that the reciprocal of this infinite series splits 
off instantaneously the complex-valued in-medium loop ${\cal L}$ and the third 
equality shows that the result is identical to a geometrical series with an 
effective density and momentum dependent potential: 
\begin{equation} V_{\rm eff} = {C_0+2C_2 \vec q^{\,2}+C_2^2(Y-X  \vec q^{\,2}) 
\over (1-C_2 X)^2 }\,. \end{equation} 
The beginning and the end of eq.(10) represent a remarkable identity. It means 
that although the individual multi-loop diagrams did not factorize, their total 
sum agrees exactly with a geometrical series one would have obtained by using the 
effective potential $V_{\rm eff}$ and assuming factorization. Note that $V_{\rm eff}$ 
depends in a non-linear way on the off-shell term $X$ and when dropping 
the effective range $C_2=0$ one recovers $V_{\rm eff}=C_0 =4\pi a/M$. A resummation 
of the in-medium T-matrix with inclusion of all partial waves has been proposed in 
section 4.2 of ref.\cite{lacour} (assuming on-shell factorization). 

The intermediate resummation result in eq.(10) does not yet provide the correct 
integrand for computing the energy per particle $\bar E(k_f)$ because it is 
complex-valued. Following the instructive diagrammatic analysis in section 4 of 
ref.\cite{resum} we make the conjecture that the proper resummation of 
in-medium ladder diagrams to all orders is achieved by the substitution:  
\begin{equation} {1\over V_{\rm eff}^{-1}- {\cal L}} \rightarrow {1\over {\rm Im} 
{\cal L}}\,\arctan{{\rm Im}  {\cal L} \over  V_{\rm eff}^{-1}- {\rm Re}{\cal L}}
\,. \end{equation} 
\begin{figure}
\begin{center}
\includegraphics[scale=0.5]{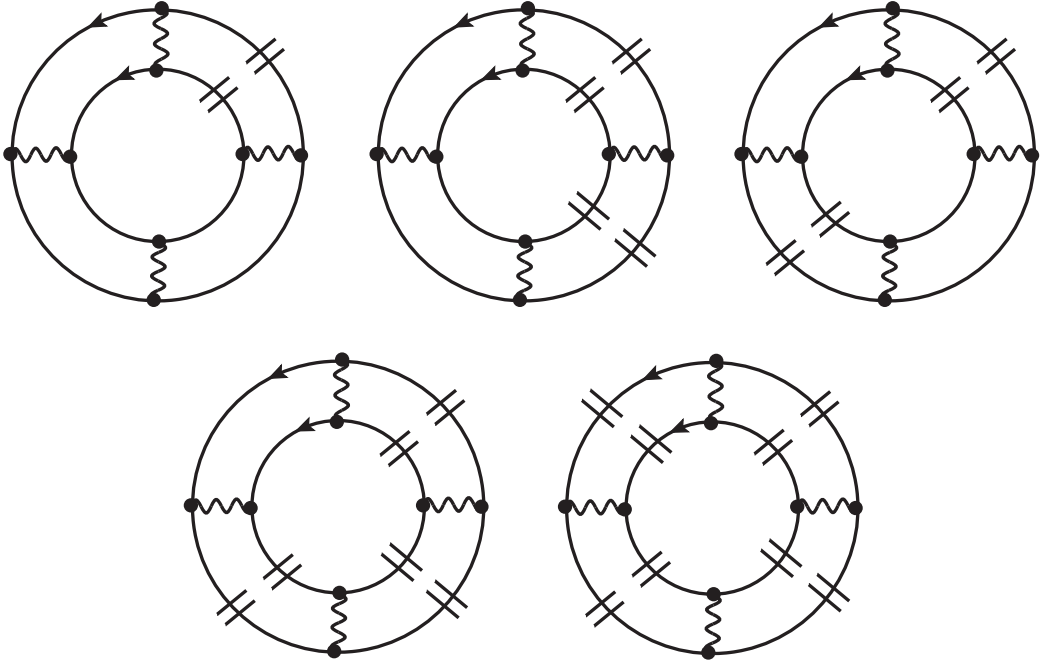}
\end{center}
\vspace{-.5cm}
\caption{In-medium ladder diagrams contributing to the energy density at fourth 
order. The third and fifth diagram have symmetry factors $1/2$ and $1/4$, 
respectively.}
\end{figure}
We have verified by a detailed diagrammatic analysis that the 
arctangent-prescription in eq.(12) leads to the correct result in a perturbative 
expansion in the coupling constants $C_{0,2}$ (at least) up to sixth order. In order 
to present an example, the corresponding set of diagrams at fourth order is shown 
in Fig.\,3. These decorated diagrams have symmetry factors $1, 1, 1/2, 1$ and 
$1/4$, in the order shown. Taking these symmetry factors into account the 
integrand for computing the energy density at fourth order has the form: 
\begin{equation} T_4^*+T_3^* T_1(2i\,{\rm Im}{\cal L})+{1\over 2} (T_2^*)^2 
(2i\,{\rm Im}{\cal L}) +T_2^* (T_1)^2 (2i\,{\rm Im}{\cal L})^2 +{1\over 4} 
(T_1)^4 (2i\,{\rm Im}{\cal L})^3 \,, \end{equation}
where $T_n^*$ denotes the complex-conjugate\,\footnote{We are using here the fact 
derived in section 3 of ref.\cite{resum} that the complex-valued in-medium 
loop ${\cal L}$ goes over into its complex-conjugate if the diagram with double 
medium insertions is taken out.} of the $n$-th order T-matrix $T_n$ to be 
calculated recursively via eq.(8). This sum of five complex terms is indeed 
real-valued and it agrees exactly with the expression one obtains by expanding 
the arctangent-function in eq.(12) to fourth power in the couplings $C_{0,2}$. 
The relevant real-valued integrands at second and third order are: $T_2^*+{1\over 2}
(T_1)^2(2i\,{\rm Im}{\cal L})$ and $T_3^*+T_2^*T_1(2i\,{\rm Im}{\cal L})+{1\over 3} 
(T_1)^3 (2i\,{\rm Im}{\cal L})^2$. Going step by step perturbatively to higher $n$ is 
in principle straightforward, it requires to check in each case the equality of 
two very long polynomial expressions. A complete verification of  
eq.(12) to all orders $n$ requires the consideration of all possible partitions 
$(n_1,n_2,\dots,n_j)$ of $n=n_1+n_2+\dots +n_j$ together with their symmetry factors. 
The sequence $(n_1,n_2,\dots,n_j)$ counts here the numbers of contact-interactions 
between consecutive double medium-insertions (see Fig.\,3). Since the partitions 
of $n$ are in one-to-one correspondence to the inequivalent irreducible 
representations of the permutation group $S_n$ there might exist some sophisticated 
combinatorial identities, which could help to solve this problem completely.

Before turning to the final result for the resummed energy per particle we give 
the explicit analytical expressions for the off-shell terms:  
\begin{equation} X = - {M k_f^3 \over 3\pi^2}\,, \qquad Y=  - {M k_f^5 \over 
15\pi^2} (5s^2+ 3)\,, \end{equation} 
and for the complex-valued in-medium loop:
\begin{equation}  {\cal L} = {M k_f \over 4\pi^2} \Big\{ - R(s,\kappa)+ 
i \pi\, I(s,\kappa)\Big\}\,. \end{equation} 
The dimensionless variables $s= |\vec p_1+\vec p_2|/2k_f$ and $\kappa= 
|\vec p_1-\vec p_2|/2k_f$ fulfil the constraint $s^2+\kappa^2 <1$, since 
$|\vec p_{1,2}|<k_f$. The dimensionless real part $R(s,\kappa)$ and imaginary part 
$I(s,\kappa)$ introduced in eq.(15) have the form \cite{resum}:
\begin{equation} R(s,\kappa) = 2 +{1\over 2s}[1-(s+\kappa)^2]\ln{1+s +\kappa
\over |1-s -\kappa|}+{1\over 2s}[1-(s-\kappa)^2]\ln{1+s -\kappa
\over 1-s +\kappa}\,, \end{equation}
\begin{equation} I(s,\kappa) = \kappa \, \theta(1-s-\kappa) + {1\over 2s} 
(1-s^2-\kappa^2)\, \theta(s+\kappa-1)\,.  \end{equation}
The same expression for the in-medium loop has been derived in appendix C 
of ref.\cite{lacour}. With these ingredients the resummed energy per particle is 
readily calculated. The 2-ring Hartree and 1-ring Fock diagram add up with their 
spin-factors as $4-2=2$. One uses the reduction formula eq.(15) in 
ref.\cite{resum} for the integral over two Fermi spheres $|\vec p_{1,2}|<k_f$ to 
cancel yet the factor $1/{\rm Im}{\cal L}$  in eq.(12), and obtains the following 
expression for the interaction energy per particle:   
\begin{equation} \bar E(k_f)= -{24k_f^2\over \pi M}\int\limits_0^1 \!ds\, s^2  
\!\!\int\limits_0^{\sqrt{1-s^2}}  \!\!d\kappa \, \kappa \, \arctan {\pi\,
I(s,\kappa) \over \Omega(a,r_s)^{-1}+ R(s,\kappa)}\,, \end{equation}
with the auxiliary function:
\begin{equation}  \Omega(a,r_s) = {3 a k_f\over(3\pi - a^2 r_s k_f^3)^2} 
\bigg\{ {3\pi \over 2}(2- a r_s k_f^2 \kappa^2) +{a^3 r_s^2 k_f^5\over 20} 
(5\kappa^2-5s^2-3)\bigg\}\,, \end{equation}
depending on the physical parameters: scattering length $a$, effective range 
$r_s$ and Fermi momentum $k_f$. When dropping the effective range one recovers 
$\Omega(a,0)=a k_f/\pi$ and in the limit of large scattering length, 
$\lim\limits_{a\to \infty} \Omega(a,r_s)= 3(5\kappa^2-5s^2-3)/20$ becomes 
independent of $r_sk_f$. This comparison demonstrates that the limits $ a\to \infty$ 
and $r_s\to 0$ do not commute. The arctangent-function occurring in eq.(18) 
refers to the usual branch with odd parity, $\arctan(-x) =-\arctan x$, 
and values in the interval $[-\pi/2,\pi/2]$. We note that the leading order correction 
involving the effective range, $\bar E(k_f)^{(\rm lo)} =  a^2 r_s k_f^5 /20\pi M$, 
agrees with the low-density expansion of ref.\cite{hammer}.  


Since the unitary limit $ a\to \infty$ can be performed straightforwardly for
eq.(19) a first interest aims at the Bertsch parameter. A numerical 
evaluation of the double-integral in eq.(18) gives $\xi_n(r_s\ne 0) =0.876$, 
which is considerably larger than the value $\xi_n(r_s= 0) =0.507$ for vanishing 
effective range. The reason for this feature are the off-shell terms $X, Y$ 
in eq.(11) which change the reciprocal effective interaction $V_{\rm eff}^{-1}$ 
drastically. Note that the ratio $C_2/C_0=-a r_s/4$ is large. The resummation 
formula given in eqs.(18,19) offers of course the possibility to study various 
other strong coupling limits $ a\to \infty$, by imposing a certain relation between 
$a$ and $r_s$.        

\subsection{Particle-particle ladders only}
In this subsection we adapt the previous methods to the particle-particle 
ladders series with inclusion of the effective range correction \cite{schaefer}. 
The particle-particle ladder diagrams are resummed via a geometrical series 
$( V_{\rm eff}^{-1}- {\cal L})^{-1}_{pp}$, in which the quantities $X, Y$ and 
${\cal L}$ take on a different form. For particle-particle intermediate states 
only the integration region in eq.(5) changes effectively from 
the sum to the union of two partly overlapping Fermi spheres $|\vec P \pm \vec 
l\,|<k_f$. Performing these integrals, one finds for the $pp$ off-shell terms:
\begin{equation} X_{pp} = - {M k_f^3 \over 12\pi^2}(1+s)^2(2-s)\,, \qquad 
Y_{pp}=  - {M k_f^5 \over 60\pi^2} (1+s)^3(6-3s+s^2)\,, \end{equation}
in the region $0<s<1$ of interest, and for the $pp$ in-medium loop:
\begin{equation}{\cal L}_{pp} = -{Mk_f \over 4\pi^2} F_{pp}(s,\kappa)\,, 
\end{equation}
with the familiar particle-particle bubble function ($s^2+\kappa^2<1)$:
\begin{equation} F_{pp}(s,\kappa) = 1+s -\kappa \ln{1+s +\kappa \over 1+s 
-\kappa}+{1\over 2s}(1-s^2-\kappa^2)\ln{(1+s)^2 -\kappa^2 \over 1-s^2 -
\kappa^2}\,. \end{equation}
Putting these pieces together, one derives from the resummed 
particle-particle ladder diagrams the following expression for the energy 
per particle:
\begin{equation} \bar E(k_f)= -{24k_f^2\over M}\int\limits_0^1 \!ds\, s^2  
\!\!\int\limits_0^{\sqrt{1-s^2}}  \!\!d\kappa \, \kappa \, {I(s,\kappa) \over 
\Omega_{pp}(a,r_s)^{-1}+ F_{pp}(s,\kappa)}\Bigg\} \,, \end{equation}
with the auxiliary function:
\begin{eqnarray} \Omega_{pp}(a,r_s) &=& {3 a k_f\over[12\pi - a^2 r_s k_f^3
(1+s)^2(2-s)]^2} \bigg\{ 24\pi (2- a r_s k_f^2 \kappa^2) \nonumber \\ &&
+a^3 r_s^2 k_f^5(1+s)^2 \bigg[(2-s) \kappa^2+{1+s\over 5}(3s-6-s^2)\bigg]
\bigg\}\,,\end{eqnarray}
depending on the physical parameters $a, r_s$ and $k_f$. Again, when dropping 
the effective range, $\Omega_{pp}(a,0)=a k_f/\pi$, and in the strong coupling 
limit $\Omega_{pp}(\infty,r_s)=3[(1+s)(3s-6-s^2)+5(2-s)\kappa^2]/[5(1+s)^2
(2-s)^2]$ becomes independent of $r_sk_f$. Because of zeros in the denominator, 
the double-integral in eq.(23) is to be understood as a principal-value integral. 
In the actual numerical treatment we work with a cutting-function, ${\rm Cut}(x) = x$ 
for $|x|<N$, ${\rm Cut}(x) = 0$ for $|x|>N$, in order to implement the symmetrical 
excision around the first-order poles and study the convergence behavior for 
large $N$. It is important to note that our result agrees exactly with eq.(13) in 
ref.\cite{schaefer} (after matching the conventions for the coupling constants $C_{0,2}$). 
The underlying structure of a geometrical series with an effective potential 
$V_{\rm eff}^{(pp)}$ and the reduction to an integral over the quarter unit-disc has 
seemingly not been recognized in that paper. Note that ref.\cite{schaefer} has 
investigated in addition the summable series of the particle-hole ring diagrams. For 
this subclass of diagrams (as well as for the isolated hole-hole ladder series) the 
energy per particle $\bar E(k_f)$ diverges in the strong coupling limit $a\to \infty$ 
(see sections 3 and 4 in ref.\cite{schaefer}).

In the unitary limit $a\to \infty$, a numerical integration gives for the
Bertsch parameter $\xi_n^{(pp)}(r_s\ne 0) \simeq 0.43$ which amounts to almost 
twice the value $\xi_n^{(pp)}(r_s= 0) \simeq 0.237$ at vanishing effective range.
The approach of Schwenk and Pethick \cite{schwenk} has employed the form 
$\Omega_{pp}(a,r_s)^{-1}=\pi(1/a k_f+r_s k_f\kappa^2/2)$ as suggested by the 
effective range expansion in free space. 
\begin{figure}
\begin{center}
\includegraphics[scale=0.5]{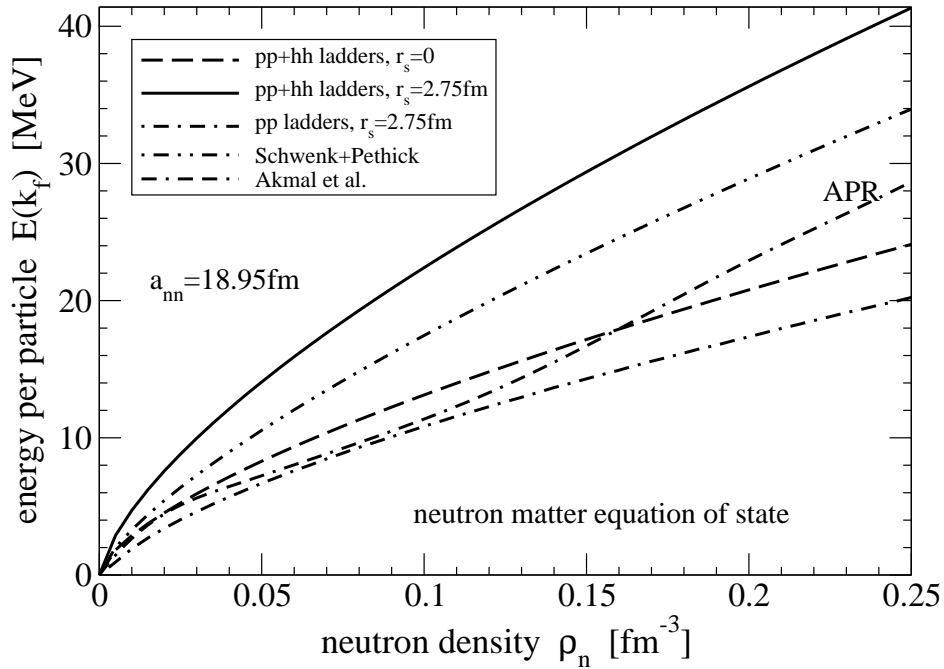}
\end{center}
\vspace{-.5cm}
\caption{Energy per particle of neutron matter as a function of the neutron 
density $\rho_n = k_f^3/3\pi^2$. The line labelled APR stems from the 
sophisticated many-body calculation of ref.\cite{akmal}. }
\end{figure}
\subsection{Application to low-density neutron matter}
As an application we consider the equation of state of neutron matter. Due to the 
very large neutron-neutron scattering length $a_{nn}=(18.95\pm 0.40)\,$fm 
\cite{chen} neutron matter at low densities is supposed to be a Fermi gas close 
to the unitary limit. Fig.\,4 shows the energy per particle of 
neutron matter as a function of the neutron density $\rho_n = k_f^3/3\pi^2$.  
The full and dashed line result from the resummation of (combined particle-hole) 
ladder diagrams eq.(18) with inclusion of the empirical effective range parameter 
$r_s=(2.75\pm 0.11)\,$fm and by dropping it, $r_s=0$. The resummation of the 
particle-particle ladders only leads to dashed-dotted line in Fig.\,4 and the 
result of the approach by Schwenk and Pethick \cite{schwenk} is additionally 
included. In each case the sum of kinetic and interaction energy per particle 
$3k_f^2/10M + \bar E(k_f)$ is plotted. The curve labeled APR stems from the 
sophisticated many-body calculation by the Urbana group \cite{akmal}, to be 
considered as a representative of realistic 
neutron matter calculations. In fact a recent calculation of neutron matter in 
chiral effective field theory with inclusion of subleading three-nucleon forces 
etc. leads to a very similar equation of state \cite{neun3lo}. Only the lower 
two lines in Fig.\,4 give a reasonable reproduction of the APR curve at 
low densities $\rho_n < 0.2\,$fm$^{-3}$. Since the dimensionless variable 
$a_{nn}k_f$ reaches values up to $34.3$ in this density region the behavior of the total
energy per particle $3k_f^2/10M +\bar E(k_f)$ is almost entirely determined by the value of 
the Bertsch parameter $\xi_n$. Note that for a correct reproduction of the neutron 
matter equation of state at low densities it has to be close to $\xi_n\simeq 0.5$.
The additional repulsive effects arising from hole-hole and mixed particle-hole 
ladders are recognizable as the difference between the full and dashed-dotted line in 
Fig.\,4. Despite the deviations from the APR curve visible in Fig.\,4, we reproduce 
the results of Schwenk and Pethick for densities $\rho_n<0.02\,$fm$^{-3}$ actually 
considered in ref.\cite{schwenk}. 

\section{Resummation of p-wave contact interaction}
In this section we extend the resummation method to a spin-independent p-wave 
contact-interaction.\footnote{Additional spin-orbit and tensor forces are 
necessary to describe the splitting of the three $^3P_{0,1,2}$ channels.} The 
pertinent contact-vertex obeying Galilei invariance reads:  
\begin{equation}  C_1\, \vec q_{\rm in} \cdot\vec q_{\rm out} \,, \end{equation}
with the coupling constant $C_1$ related to the p-wave scattering volume $a_1^3$ 
by $C_1 = 12\pi a_1^3/M$. The repeated rescatterings via this contact-vertex in 
the vacuum (i.e. the left diagram in Fig.\,2 iterated to all orders) leads to a 
p-wave phase shift of the form:
\begin{equation} \tan \delta_1(q) = (a_1 q)^3\,,  \end{equation}
with $q$ the center-of-mass momentum and we have used the rule 
$\int_0^\infty dl\,l^n=0$ of dimensional regularization. In the in-medium loop 
two consecutive $C_1$ vertices introduce the factor $(\vec q_{\rm in} \cdot \vec 
l\,)(\vec l\cdot \vec q_{\rm out})$. We decompose the emerging tensorial integral 
over $l_i l_j $ into the transversal projector $\delta_{ij}- \hat P_i \hat P_j$ 
and the longitudinal projector  $\hat P_i \hat P_j$, where $\vec P = (\vec p_1+ 
\vec p_2)/2$ (see Fig.\,2). Since these are two orthogonal projectors the 
factorization of multi-loop diagrams holds separately for the transversal and 
longitudinal part of the tensor to be contracted with $q_{\rm in}^i q_{\rm out}^j$.
In this decoupling procedure one encounters a transversal in-medium loop whose  
real and imaginary part read: 
\begin{eqnarray} R_\perp(s,\kappa) &=& {2\over 3}-{s^2\over 4}+2\kappa^2+
{(1-\kappa^2)^2\over 4s^2}  +{[(s+\kappa)^2-1]^2\over 16s^3}(s^2+\kappa^2
-4s\kappa-1 )\nonumber \\ && \times \ln{1+s +\kappa \over |1-s -\kappa|} 
+{[(s-\kappa)^2-1]^2\over 16s^3}(s^2+\kappa^2+4s\kappa -1 )\ln{1+s -\kappa 
\over 1-s +\kappa}\,,\end{eqnarray}
\begin{equation} I_\perp(s,\kappa) = \kappa^3 \, \theta(1-s-\kappa) + 
{1-s^2-\kappa^2 \over 16s^3}\Big[12 s^2\kappa^2-(1-s^2-\kappa^2)^2\Big]\,
\theta(s+\kappa-1)\,,  \end{equation}
as well as a longitudinal in-medium loop whose real and imaginary part read: 
\begin{eqnarray} R_\parallel(s,\kappa) &=& {8\over 3}+{s^2\over 2}+2\kappa^2
-{(1-\kappa^2)^2\over 2s^2}  +\bigg[{(1-s^2-\kappa^2)^3 \over 8s^3}-\kappa^3
\bigg] \nonumber \\ && \times \ln{1+s +\kappa \over |1-s -\kappa|} +\bigg[
{(1-s^2-\kappa^2)^3 \over 8s^3}+\kappa^3 \bigg] \ln{1+s -\kappa \over 1-s 
+\kappa}\,,\end{eqnarray}
\begin{equation} I_\parallel(s,\kappa) = \kappa^3 \, \theta(1-s-\kappa) + 
{(1-s^2-\kappa^2)^3 \over 8s^3} \,\theta(s+\kappa-1)\,.  \end{equation}
The diagrammatic analysis in section 4 of ref.\cite{resum} emphasizing the important 
role of symmetry factors instructs that for the energy density the proper resummation 
to all orders is achieved through an arctangent-function. By using the
operator identity for projectors, $\arctan(\Pi_\perp x_\perp+\Pi_\parallel x_\parallel)
= \Pi_\perp\arctan x_\perp+\Pi_\parallel \arctan x_\parallel$, the transversal and 
longitudinal parts get separated completely. A special feature of the
(resummed) p-wave contact-interaction is that the 2-ring Hartree and 1-ring Fock 
diagram add up with their spin-factors as $4+2=6$. While $\vec q_{\rm in}=(\vec p_1-
\vec p_2)/2=\vec q$ for both, there is a sign difference in $\vec q_{\rm out}=\pm \,
\vec q$ for the Hartree and Fock contributions. Putting all pieces together one finds 
from the resummed ladder diagrams with a (spin-independent) p-wave 
contact-interaction the following expression for the energy per particle:
\begin{eqnarray} \bar E(k_f) &=& -{72k_f^2 \over \pi M} \int\limits_0^1 \!ds\, 
s^2  \!\!\int \limits_0^{\sqrt{1-s^2}}  \!\!d\kappa \, \kappa \, \Bigg\{2\arctan 
{I_\perp(s,\kappa) \over (a_1k_f)^{-3}+ \pi^{-1}R_\perp(s,\kappa)} \nonumber \\ 
&& \qquad \qquad \qquad \qquad \qquad \,\,+\arctan {I_\parallel(s,\kappa) \over 
(a_1k_f)^{-3}+ \pi^{-1}R_\parallel(s,\kappa)}  \Bigg\}\,. \end{eqnarray} 
The factor $2$ is reminiscent of two transversal versus one longitudinal 
degree of freedom. Another interesting feature is that the imaginary parts of the 
transversal and longitudinal loop functions, $I_\perp(s,\kappa)$ and $I_\parallel
(s,\kappa)$ in eqs.(28,30), appear (with a prefactor $s^2\kappa$) as weighting 
functions in the reduction of integrals over two Fermi spheres $|\vec p_{1,2}|<k_f$.

As a first application let us consider the low-density expansion of 
$\bar E(k_f)$ in eq.(31), which reads:
\begin{equation} \bar E(k_f) = {k_f^2 \over 2M}\bigg\{ -{9 \over 5\pi}
(a_1k_f)^3+{1033-156\ln2 \over 385 \pi^2}(a_1k_f)^6 + \dots \bigg\}\,\,.  
\end{equation}
The linear term in the scattering volume $a_1^3$ agrees with ref.\cite{hammer} 
(after matching conventions $a_p = -3^{1/3} a_1 $) while the second order term 
with its peculiar numerical coefficient is new. It is also straightforward to 
perform the strong coupling limit $a_1\to \infty$ for $\bar E(k_f)$ in eq.(31). 
One finds $\bar E(k_f)^{(\infty)}= (3k_f^2/10M)(-1.948)$, corresponding to almost 
minus twice the free Fermi gas energy and therefore an instability of the system. 
The basic reason for this overly strong p-wave attraction is the enhancement factor 
$3$ from the coherent sum of Hartree and Fock terms. In the case of a strong s-wave 
interaction the number multiplying the free Fermi gas energy $3k_f^2/10M$ is 
$\xi_n-1 = -0.493 \simeq -1.948 \cdot 1/4$.  
\subsection{Particle-particle ladders only}
In this subsection we study for comparison the resummation of only 
particle-particle ladders with a (spin-independent) p-wave contact-interaction. 
The factorization of multi-loop diagrams and the separation into transversal and 
longitudinal parts hold in the same way, while the resummation to all orders is 
accomplished now by a geometrical series. The resulting expression for the 
energy per particle reads:     
\begin{equation} \bar E(k_f) = -{72k_f^2 \over M} \int\limits_0^1 \!ds\, s^2  
\!\!\int \limits_0^{\sqrt{1-s^2}}  \!\!d\kappa \, \kappa \, \Bigg\{ {2I_\perp
(s,\kappa) \over \pi(a_1k_f)^{-3}+ F_{pp}^\perp(s,\kappa)} +{I_\parallel(s,\kappa) 
\over \pi (a_1k_f)^{-3}+ F_{pp}^\parallel(s,\kappa)} \Bigg\} \,, \end{equation} 
supplemented by a principal-value prescription for the double-integral. The transversal 
and longitudinal particle-particle bubble functions $F_{pp}^{\perp,\parallel}(s,\kappa)$ 
occurring in the denominators have the form:
\begin{eqnarray} F_{pp}^\perp(s,\kappa)&=& {1\over 3}+{(1-\kappa^2)^2\over 8s^2} 
+\kappa^2 -{s^2 \over 8}-{5s^3 \over 12}+{s\over 8}(6+7\kappa^2) \nonumber\\ &&
-{\kappa^2 \over 8s}(1-\kappa^2) -\kappa^3 \ln{1+s+\kappa \over 1+s-\kappa}
+{1-s^2-\kappa^2 \over 16 s^3} \nonumber \\ && \times \Big[2s^2(1+5\kappa^2)-s^4-
(1-\kappa^2)^2\Big]\ln{(1+s)^2-\kappa^2\over 1-s^2-\kappa^2 }\,,\end{eqnarray} 
\begin{eqnarray} F_{pp}^\parallel(s,\kappa)&=& {4\over 3}- {(1-\kappa^2)^2 \over 
4s^2}+\kappa^2+{s^2 \over 4}-{s^3\over 6}+{s\over 4}(6+5\kappa^2) +{\kappa^2 
\over 4s}(1-\kappa^2) \nonumber\\ &&-\kappa^3 \ln{1+s+\kappa \over 1+s-\kappa}
+{(1-s^2-\kappa^2)^3 \over 8 s^3} \ln{(1+s)^2-\kappa^2\over 1-s^2-\kappa^2 }\,.
\end{eqnarray}
These functions satisfy the relations $F_{pp}^{\perp,\parallel}(s,\kappa)+ {\rm Re}\,
F_{pp}^{\perp,\parallel} (-s,\kappa)=R_{\perp,\parallel}(s,\kappa)$ inside the region 
$s^2+\kappa^2<1$ of interest. As it must be the low-density expansion of 
$\bar E(k_f)$ in eq.(33) agrees up to next-to-leading order with eq.(32). The 
additional hole-hole and mixed particle-hole ladders included in eq.(31) are of 
cubic and higher order in the scattering volume $a_1^3$. The strong coupling 
limit $a_1\to \infty$ is also readily performed. One gets $\bar 
E(k_f)^{(\infty)}= (3k_f^2/10M)(-2.405)$, which amounts to about $23\%$ more 
attraction than in the case of the complete ladder series. It is a general feature that  
hole-hole and mixed particle-hole ladders generate some moderate repulsion. Again the 
enhancement factor 3 in comparison to the case of a strong s-wave interaction is 
recognizable here, $\xi_n^{(pp)}-1 \simeq -0.763 \simeq -2.405 \cdot 1/3$. 

\begin{figure}
\begin{center}
\includegraphics[scale=0.55]{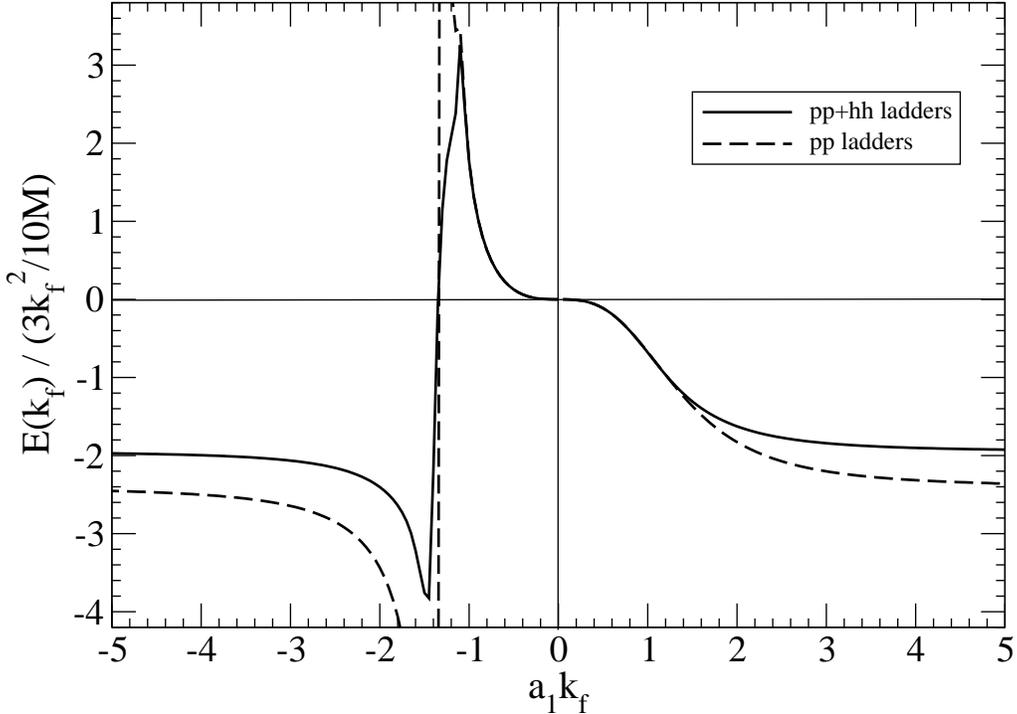}
\end{center}
\vspace{-.4cm}
\caption{Energy per particle $\bar E(k_f)$ from p-wave ladder diagrams divided 
by the free Fermi gas energy $3k_f^2/10M$ as a function of the dimensionless parameter 
$a_1k_f$.}
\end{figure}

It is interesting to explore the dependences on density and coupling strength 
of the novel p-wave resummation results eqs.(31,33). For this purpose we plot in 
Fig.\,5 the ratio of the interaction energy $\bar E(k_f)$ to the free Fermi gas 
energy $3k_f^2/10M$ as a function of the dimensionless parameter $a_1k_f$. The 
full and dashed curve correspond to the complete ladder series and partial 
$pp$-ladder series, respectively. On the attractive side  $a_1k_f>0$ one observes 
that both curves stay closely together up to $a_1 k_f \simeq 1.5$ and from 
there upwards the $pp$-ladder series develops somewhat more attraction. The 
behavior on the repulsive side $a_1k_f<0$ is characterized by a sharp peak of 
the full curve at $a_1 k_f\simeq -1.1$ and an even higher peak of the dashed 
curve at $a_1 k_f\simeq -1.3$. Beyond that a non-perturbative behavior sets in, 
such that both curves pass through zero at $a_1 k_f\simeq -1.35$ and attraction 
gets generated from the repulsive scattering volume $a_1^3<0$. In order to join
smoothly with the attractive side at infinity $a_1 k_f=\pm \infty$, this attraction 
is more strongly pronounced for the $pp$-ladder series. 

As a side remark we note that s-wave and p-wave contact-interactions do not 
interfere in the ladder diagrams. The usual argument of rotational invariance 
(in momentum-space) does not apply in the medium. An sp-interference term 
introduces into the in-medium loop a factor that is odd under parity $\vec l \to 
-\vec l$, but the integration regions defined by two overlapping or intersecting 
Fermi spheres in eqs.(5,6) are even under this transformation, and therefore it 
integrates always to zero.

Let us add a final remark on the physical applicability of the resummation results 
derived in this work. In order to obtain the factorization of multi-loop diagrams
(in a generalized sense) the input interactions have to be contact-interactions.
Their iteration to all orders in the vacuum leads to a unitary S-matrix with 
phase-shifts given by eqs.(4,26). On the other hand the Wigner bound 
\cite{wignerbound} due causality states that a scattering phase-shift must not 
drop too fast with momentum, $\partial \delta_l(q)/ \partial q>- \bar R$, where 
$\bar R$ measures the range of the interaction. In the case of (unresolved) 
contact-interactions the range $\bar R\simeq 0$ vanishes practically. This  
consideration could provide an argument for the condition that the resummation 
results are physically applicable only for positive scattering length $a>0$, 
positive scattering volume $a_1>0$, and negative effective range $r_s<0$.
\section*{Appendix: Resummation with coupled channels} 
In this appendix we supply the result for the complete resummation of in-medium 
ladder diagrams due to an s-wave contact-interaction in a binary (i.e. two-component)
many-fermion system. The prototype of such a system is nuclear matter made up from 
strongly interacting protons and neutrons. Low-energy nucleon-nucleon scattering 
is characterized by two different scattering lengths, the spin-singlet $(I=1)$ 
scattering length $a_s\simeq  19\,$fm and the spin-triplet $(I=0)$ scattering 
length $a_t \simeq -5.4\,$fm, where $I=0,1$ denotes the total isospin. The 
many-body dynamics is richer and offers the possibility to introduce an 
isospin-asymmetry through different (variable) proton and neutron densities (or 
different Fermi momenta $k_p$ and $k_n$). Returning to the in-medium loop in Fig.\,2 
one sees that one has to deal now with four (partly coupled) states: $(pp, nn, 
pn, np)$. The last two channels get decoupled through a $45^\circ$ rotation to the 
isospin-eigenstates $(pn\pm np)/\sqrt{2}$. In this isospin-basis the 
momentum-independent contact-interaction (proportional to the scattering lengths 
$a_{s,t}$) is diagonal and therefore the multi-loop diagrams factorize. As a new 
element one encounters mixed proton-neutron loops with integration regions defined by 
eqs.(5,6), where one Fermi momentum $k_f$ is $k_p$ and the other is $k_n$. The real 
part of this mixed in-medium loop is given simply by the sum of two independent proton 
and neutron terms, but a more complicated overlap of two Fermi-spheres $|\vec P\pm 
\vec l\,|<k_{p,n}$ and the surface  $|\vec l\,| = |\vec q\,|$ has to be evaluated for the 
imaginary part (see eq.(38) below). The resummation of the particle-hole ladder 
diagrams to all orders through an arctangent-function is also easily carried out, since 
for a $4\times 4$ diagonal matrix one has the identity: $\arctan({\rm diag}(d_1,d_2,d_3,
d_4))={\rm diag}(\arctan d_1, \arctan d_2, \arctan d_3, \arctan d_4)$. In order to 
compute the energy density one rotates back to the particle basis and adds the Hartree 
and Fock contributions with their spin-factors $4$ and $-2$. Putting all pieces 
together the final result for the energy per particle reads:  
\begin{eqnarray} \bar E(k_p,k_n) &\!\!\!\!=\!\!\!\!& -{24 \over \pi M(k_p^3+k_n^3)} 
\Bigg\{\int\limits_0^1 \!ds\, s^2  \!\!\int\limits_0^{\sqrt{1-s^2}}  \!\!d\kappa \, 
\kappa \, \Bigg[k_p^5\arctan {I(s,\kappa) \over (a_sk_p)^{-1}+ \pi^{-1}R(s,\kappa)} 
\nonumber \\ && + \,k_n^5 \arctan {I(s,\kappa)\over(a_sk_n)^{-1}+\pi^{-1}R(s,\kappa)}
\Bigg]+\int\limits_0^{(k_p+k_n)/2} \!dP\, P^2\!\!\int\limits_{q_{\rm min}}^{q_{\rm max}}
\!\!dq \,q\nonumber \\ && \times \Bigg[\arctan{\Phi(P,q,k_p,k_n) \over a_s^{-1} +
(2\pi)^{-1}(k_p R_p+ k_n R_n) } +3 \arctan{\Phi(P,q,k_p,k_n)\over a_t^{-1} 
+(2\pi)^{-1}(k_p R_p+ k_n R_n) } \Bigg]  \Bigg\} \,,\nonumber \\ \end{eqnarray}
with the abbreviations $R_p = R(P/k_p,q/k_p)$ and $R_n = R(P/k_n,q/k_n)$, where 
$R(s,\kappa)$ and $I(s,\kappa)$ are given in eqs.(16,17). The integration boundaries 
are:

\begin{figure}
\begin{center}
\includegraphics[scale=0.55,clip]{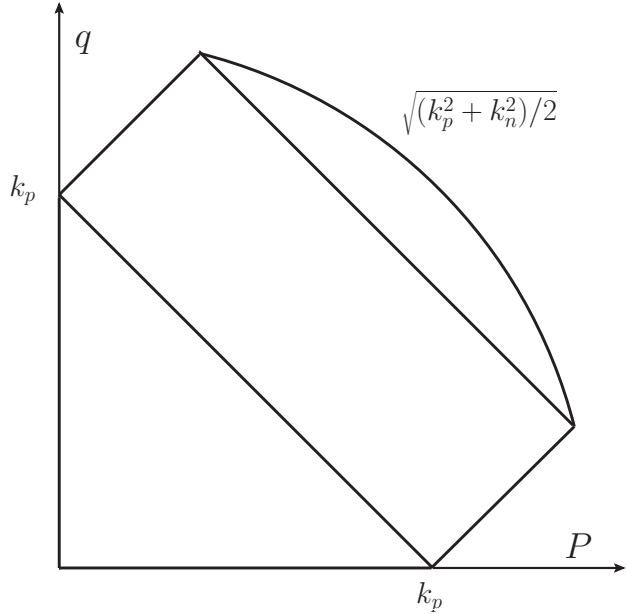}
\end{center}
\vspace{-.8cm}
\caption{Support of the function $\Phi(P,q,k_p,k_n)$ in the $(P,q)$-plane pieced 
together by a triangle, a rectangle and a circular wedge. The rectangle has 
side-lengths $\sqrt{2}\,k_p$ and $(k_n-k_p)/\sqrt{2}$.} 
\end{figure}

\begin{equation} q_{\rm min}= {\rm max}\big(0, P-k_p\big)\,,  \qquad q_{\rm max}= 
{\rm min}\Big( k_p+P, \sqrt{(k_n^2+k_p^2)/2-P^2}\,\Big)\,, \end{equation}
assuming for definiteness $k_n \geq k_p$ (i.e. an excess of neutrons over protons). 
The total nucleon density is $\rho = \rho_p+\rho_n = (k_p^3+k_n^3)/3\pi^2$. The 
imaginary part function $\Phi(P,q,k_p,k_n)$ appearing in the numerator of the last 
two $\arctan$-terms is defined piecewise by: 
\begin{eqnarray}&& \Phi =  q\,,\,\,\, {\rm for}\,\,\, P+q<k_p\,, \nonumber \\ && 
\Phi = {1\over 4P}\Big[k_p^2-(P-q)^2 \Big]\,,\,\,\, {\rm for}\,\,\,k_p<P+q<k_n\,,\, 
|P-q|<k_p\,, \nonumber \\ && \Phi= {1\over 4P}\Big[k_p^2+k_n^2-2(P^2+q^2)\Big]\,, 
\,\,\, {\rm for}\,\,\, P+q>k_n\,,\, P^2+q^2 < (k_p^2+k_n^2)/2\,. \end{eqnarray}
As illustrated in Fig.\,6 the support of this function consists of a triangle, a 
rectangle and a circular wedge pieced together. Note that $P^2 q\, \Phi(P,q,k_p,k_n)$ 
plays also the role of a weighting function (symmetric under $P\leftrightarrow q$) in 
reducing the integral over two Fermi 
spheres $|\vec p_{1,2}|<k_{p,n}$. The same expression for the $\Phi$-function as in eq.(38) 
has been derived in appendix C of ref.\cite{lacour}. The numerical implications of eq.(36) 
for the nuclear matter equation of state are discussed elsewhere \cite{schultess}. The 
inclusion of corrections from the singlet and triplet effective ranges should also be 
obvious. The inverse scattering lengths  $a_{s,t}^{-1}$ in the denominators of the 
arctan-terms in eq.(36) get then replaced by appropriate reciprocal $\Omega$-functions. 
Note also the factor $3$ in front of the last arctan-term, it ensures that the result 
$\bar E(k_f,k_f)$ for the isospin-symmetric system, $k_p=k_n=k_f$, becomes invariant under 
the interchange ($a_s \leftrightarrow a_t$) of both scattering lengths.       
As a byproduct of eq.(36) we consider the isospin-asymmetry energy $A(k_f)$. It is 
obtained by setting the Fermi momenta to $k_{n,p}=k_f(1\pm \delta)^{1/3}$ and expanding 
to quadratic order, $\bar E(k_p,k_n) = \bar E(k_f)+ \delta^2  A(k_f)+{\cal O}(\delta^4)$, 
in the isospin-asymmetry parameter $\delta$. One finds up to second order in the scattering 
lengths $a_{s,t}$:
\begin{equation} A(k_f) = {k_f^2 \over 6M} \bigg\{ 1 + {k_f \over \pi} (3a_t-a_s) 
+{8 k_f^2 \over 5\pi^2} \Big[ a_s^2(3-\ln 2)-a_t^2(2+\ln 2)\Big] + \dots \bigg\}\,,
\, \end{equation}
where the first term $k_f^2/6M$ arising from the kinetic energy has been included for 
orientation. This result is implicitly contained in ref.\cite{shortrange} via the 
parameterizations $a_{s,t}= g_A^2 M\gamma_{1,0}/(8\pi f_\pi^2)$ of the scattering 
lengths. As a special case of an asymmetric system let us finally consider the 
one-component (unitary) Fermi gas in the spin-unbalanced configuration. With one 
singlet scattering length $a$ there is interaction taking place only in the mixed 
spin-state $(\uparrow\downarrow-\downarrow\uparrow)/\sqrt{2}$ and the energy per 
particle reads: 
\begin{equation} \bar E(k_\uparrow,k_\downarrow) =  -{48 \over \pi M(k_\uparrow^3+
k_\downarrow^3)}\!\!\!\!\int\limits_0^{(k_\uparrow+k_\downarrow)/2} \!\!\!dP\, P^2  \!\!\int
\limits_{q_{\rm min}}^{q_{\rm max}}\!\!dq \,q \arctan{\Phi(P,q,k_\uparrow,k_\downarrow) 
\over a^{-1} +(2\pi)^{-1}(k_\uparrow R_\uparrow+ k_\downarrow R_\downarrow) }  \,, 
\end{equation}

\begin{figure}
\begin{center}
\includegraphics[scale=0.55,clip]{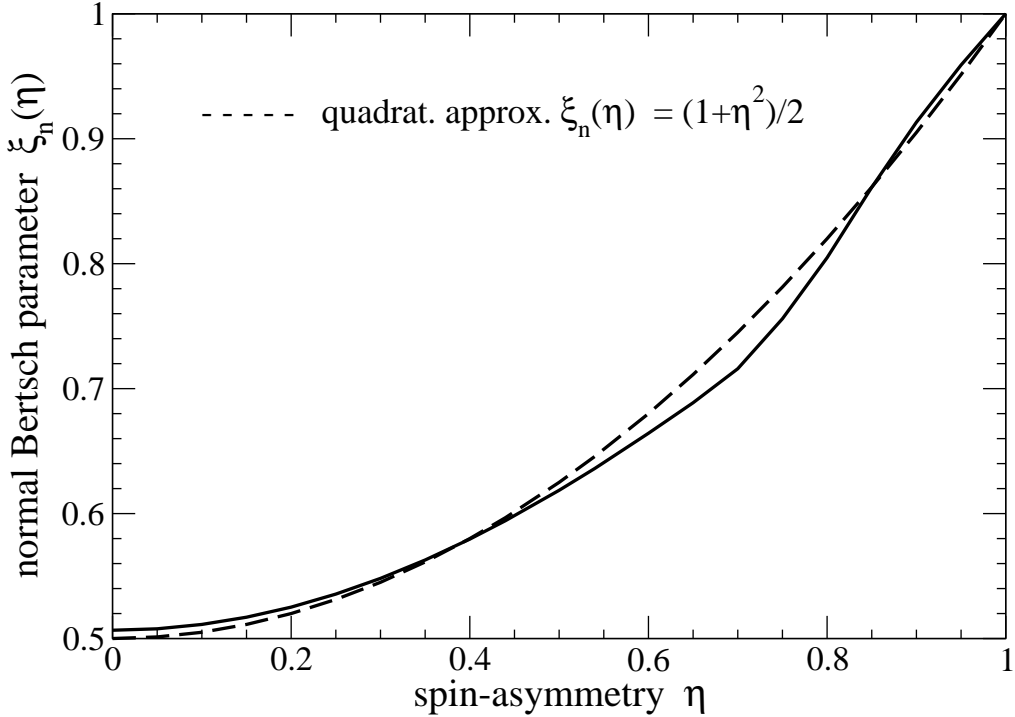}
\end{center}
\vspace{-.8cm}
\caption{The (normal) Bertsch parameter $\xi_n(\eta)$ of the unitary Fermi gas as 
a function of the spin-asymmetry  parameter $\eta$.}
\end{figure}

\noindent
with $q_{\rm max}, q_{\rm min}, R_\uparrow$, $R_\downarrow$ and 
$\Phi(P,q,k_\uparrow,k_\downarrow)$ defined in complete analogy to the proton-neutron 
system before. Note that the density of the spin-unbalanced Fermi gas is  
$\rho= (k^3_\uparrow+k^3_\downarrow)/6\pi^2$. We can use eq.(40) to determine, in the 
unitary limit $a\to \infty$, the dependence  of the (normal) Bertsch parameter 
$\xi_n(\eta)$ on the spin-asymmetry $\eta$ of the spin-unbalanced system. Setting 
the Fermi momenta of the spin-up and spin-down fermions to $k_{\uparrow,\downarrow}= k_f 
(1\pm \eta)^{1/3}$,  the quantity $\xi_n(\eta)-1$ is given by the ratio of the interaction 
energy $\bar E(k_\uparrow,k_\downarrow)$ to the kinetic energy $3k_f^2[(1+\eta)^{5/3}+
(1-\eta)^{5/3}]/20M$. The full line in Fig.\,7 shows the (normal) Bertsch parameter 
$\xi_n(\eta)$ as a function of the  spin-asymmetry parameter $\eta$. The 
result is quite close to the parabolic approximation $\xi_n(\eta)=(1+\eta^2)/2$ connecting 
the two boundary points $(\eta,\xi_n)=(0,1/2)$ and $(\eta,\xi_n)=(1,1)$.

\section*{Acknowledgements}
I thank S. Fiorilla for checking the analytical calculations underlying this work. 
Informative discussions with J.W. Holt, L. Platter, R. Schmidt, W. Weise and W. Zwerger 
are gratefully acknowledged.

\end{document}